\documentclass[superscriptaddress,twocolumn,preprintnumbers,showpacs,amsmath,amssymb]{revtex4-1}
\usepackage{graphicx}
\usepackage{dcolumn,bm}

\begin{document}

\title{Hybridization gap formation in the Kondo insulator YbB$_{12}$ observed
using time-resolved photoemission spectroscopy}

\author{M.~Okawa}
\affiliation{Department of Applied Physics,
Tokyo University of Science, Katsushika, Tokyo 125-8585, Japan}
\affiliation{Institute for Solid State Physics, University of Tokyo,
Kashiwa, Chiba 277-8581, Japan}

\author{Y.~Ishida}
\affiliation{Institute for Solid State Physics, University of Tokyo,
Kashiwa, Chiba 277-8581, Japan}

\author{M.~Takahashi}
\altaffiliation[Present address: ]{Department of Physics, University of Tokyo,
Kashiwa, Chiba 277-8561, Japan}
\affiliation{Department of Applied Physics,
Tokyo University of Science, Katsushika, Tokyo 125-8585, Japan}

\author{T.~Shimada}
\affiliation{Institute for Solid State Physics, University of Tokyo,
Kashiwa, Chiba 277-8581, Japan}

\author{F.~Iga}
\affiliation{College of Science, Ibaraki University,
Mito, Ibaraki 310-8512, Japan}
\affiliation{Department of Quantum Matter and Institute for Advanced
Materials Research, Hiroshima University, Higashi-Hiroshima 739-8530, Japan}

\author{T.~Takabatake}
\affiliation{Department of Quantum Matter and Institute for Advanced
Materials Research, Hiroshima University, Higashi-Hiroshima 739-8530, Japan}

\author{T.~Saitoh}
\affiliation{Department of Applied Physics,
Tokyo University of Science, Katsushika, Tokyo 125-8585, Japan}

\author{S.~Shin}
\affiliation{Institute for Solid State Physics, University of Tokyo,
Kashiwa, Chiba 277-8581, Japan}
\affiliation{CREST, Japan Science and Technology Agency, Chiyoda, Tokyo 102-0075, Japan}

\date{\today}

\begin{abstract}
A detailed low-energy electronic structure of a Kondo insulator YbB$_{12}$ was revealed by a synergetic
combination of ultrahigh-resolution laser photoemission spectroscopy (PES) and time-resolved PES.
The former confirmed a 25-meV pseudogap corresponding to the Kondo temperature of this
material, and more importantly, it revealed that a 15-meV gap and a Kondo-peak
feature developed below a crossover temperature $T^\ast \sim 110$ K.
In harmony with this, the latter discovered a very long recombination time exceeding 100 ps
below $\sim$$T^\ast$.
This is a clear manifestation of photoexcited carriers due to the bottleneck in the recovery dynamics,
which is interpreted as a developing hybridization gap of a hard gap.
\end{abstract}


\preprint{Journal-ref.: Phys.\ Rev.\ B \textbf{92}, 161108(R) (2015)}

\maketitle

% \section{Introduction}

Due to the strong electron correlation, lanthanide and actinide $f$ electron systems exhibit various
ground states such as the Fermi liquid metal with valence fluctuation, magnetic order, quantum criticality,
and exotic superconductivity.
Among them, there is a class of rare-earth compounds called the Kondo insulators/semiconductors that show a
metal-to-insulator crossover accompanied by a loss of local magnetic moment \cite{Takabatake98,*Riseborough00}.
YbB$_{12}$ is a typical Kondo insulator \cite{Kasaya83,Iga98,Iga99} with a fluctuating Yb valence
around 2.9+ \cite{Takeda04,Yamaguchi09}.
According to infrared spectroscopy studies \cite{Okamura98,*Okamura05}, the Drude
response is sharply decreased below $\sim$80 K and an indirect gap of $\sim$15 meV opens at 8 K.
Inelastic neutron scattering studies reported the spin gap of 15 meV in the insulating state \cite{Alekseev04,Nemkovski07}.
This crossover is frequently explained by a narrow-gap formation due to hybridization between the conduction
band electrons and the localized $f$ electrons ($c$-$f$ hybridization) within the scheme of the periodic Anderson
model.
However, there is increasing evidence that some metallic states persist within the gap of many Kondo
insulators, making elusive whether the low-temperature ($T$) gap is a real charge gap or a pseudogap.
For example, the resistivity is still as low as $\sim$1 $\Omega\,\text{cm}$ at low $T$
in high-quality single crystals of some Kondo insulators, such as YbB$_{12}$ \cite{Iga98} and
SmB$_6$ \cite{Cooley95}.
Photoemission spectroscopy (PES) studies \cite{Susaki96,Susaki99,Susaki01,Takeda06,Yamaguchi13} reported
a pseudogap feature ($<$20 meV) that can be interpreted as a hybridization gap, but the finite spectral weight
still remained around the Fermi level $E_F$.
In addition, tunneling spectra in YbB$_{12}$ exhibit that a finite state persists at the zero-bias voltage \cite{Ekino99,Batkova06}.

In this Rapid Communication, to settle the above controversy, we investigate the electronic states within the pseudogap as well as
its evolution on cooling by an unprecedented energy resolution set to 1 meV using ultra-high-resolution laser PES
\cite{Kiss05,*Kiss08}.
Furthermore, we employed time-resolved PES (TrPES) with a pump-probe method and investigate
the nonequilibrium electron dynamics of YbB$_{12}$ at various temperatures.
The recovery time from photoexcited nonequilibrium states is the measure of a charge gap: If there is a gap as in 
insulators, semiconductors, or nodeless superconductors, the electronic recovery time will become exceedingly
long in comparison with a typical metal \cite{Rothwarf67,*Demsar03,*Demsar06}.
Such behavior was reported in past TrPES studies on typical semiconducting materials, such as GaAs and Si \cite{Bokor89}.
Since TrPES provides direct information of the electron-hole recombination, it is particularly appropriate to
investigate such dynamics.

% \section{Methods}

Single crystals of YbB$_{12}$ were grown by the floating zone method \cite{Iga98,*Iga99}.
In both ultra-high-resolution laser PES and TrPES measurements, the samples were fractured \textit{in situ}
under the base pressure of $\sim$2$\times$10$^{-11}$ Torr, and the spectra were recorded using
a VG Scienta R4000 analyzer.
In ultra-high-resolution laser PES, the incident light was the sixth harmonic (6.994 eV) of a Nd:YVO$_4$
quasi-cw laser (Spectra-Physics Vanguard) \cite{Kiss05,*Kiss08}, and the energy resolution was set to $\Delta E = 1$--2 meV.
In TrPES, the samples were excited by a $\sim$170-fs pump pulse of $h\nu_1 = 1.47$ eV at
a repetition rate of 250 kHz delivered from a Ti:sapphire amplifier (Coherent RegA 9000), and the transient was
probed by $h\nu_4 = 5.88$ eV, which is the fourth harmonic of $h\nu_1$ generated by two $\beta$-BaB$_2$O$_4$
nonlinear optical crystals \cite{Ishida11,*Ishida14}.
To strictly avoid the multiphoton photoelectrons which start to appear at the
pump fluence of  $\geq$20 $\mu$J/cm$^2$, we set it to $\sim$15 $\mu$J/cm$^2$,
which induces surface heating of $\sim$10 K at most, throughout the TrPES measurements.
The energy and time resolutions were 12 meV and 0.41 ps, respectively, and the origin of pump-probe delay 
$t = 0$ was calibrated \textit{in situ} using TrPES of graphite attached next to the sample \cite{Ishida11}.
$E_F$ was calibrated using the Fermi cutoff of evaporated Au.
At 6 and 35 K, we observed pump-induced surface photovoltages (SPVs) of 6.0 and 1.3 meV, respectively;
We corrected the rigid photovoltaic shift of the TrPES spectra due to the SPV effect at each temperature \cite{footnote}.

% \section{Results and Discussions}
% \subsection{7-eV laser PES}

\begin{figure}
\begin{center}
\includegraphics[width=80mm]{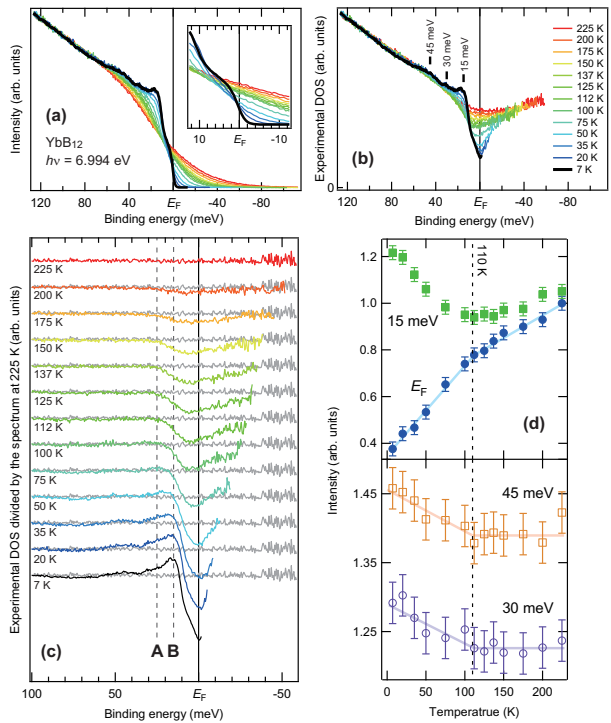}
\end{center}
\caption{
Temperature dependence of the valence band spectra of YbB$_{12}$ in the near-$E_F$ region recorded
by a 7-eV laser.
(a) Near-$E_F$ valence-band spectra at various temperatures.
The inset shows a magnified view at $E_\text{F}$.
(b) Experimental density of states (DOS).
(c) Intensity of the experimental DOS relative to that at 225 K.
(d) The spectral intensity at $E_\text{F}$, 15, 30, and 45 meV as functions of temperature.
Solid lines are guides to the eye.
Here, although the lowest temperature was in the range of 7 to 8 K, we describe it as 7 K for simplicity.
}
\label{fig1}
\end{figure}

Figure \ref{fig1}(a) shows temperature dependence of the ultra-high-resolution laser-PES spectra
of YbB$_{12}$.
These spectra were normalized to the area at $110 \pm 10$ meV.
We also show the experimental DOS in Fig.\ \ref{fig1}(b), which was obtained by dividing
the laser-PES spectra by a Gaussian-broadened Fermi-Dirac distribution function.
The spectral weight at $E_F$ gradually decreases on cooling, reflecting the opening of a pseudogap.
It is also apparent from the inset of Fig.\ \ref{fig1}(a) that the Fermi edge is present even at
the lowest temperature, which is clear evidence of in-gap states coexisting with the pseudogap.
A sharp peak at 15 meV and a broad peak at 45 meV were observed in the previous PES
studies \cite{Takeda06,Yamaguchi13}.
The former corresponds to the renormalized band by the $c$-$f$ hybridization,
whereas the latter 45-meV structure corresponds to the main peak of the Yb $4f_{7/2}$ state, which can be 
found more clearly at 30--50 meV in hard x-ray PES spectra \cite{Yamaguchi09} because of a higher
photoionization cross-sectional ratio \cite{Yeh85}.
In addition, we also find another small peak at 30 meV.
Such fine structures observed here could be considered as the crystal-field splitting
of the Yb $4f$ state \cite{Alekseev04}.

To investigate how the pseudogap and the peak develop in detail, we plot the normalized experimental DOS
with respect to the 225-K DOS in Fig.\ \ref{fig1}(c).
We can observe different evolutions of the two gap like features that are
the large pseudogap at 25 meV [A in Fig.\ \ref{fig1}(c)] and the hybridization gap at 15 meV [B in
Fig.\ \ref{fig1}(c)] on cooling;
the large gap starts opening already at 200 K, the size of which is comparable with the Kondo temperature
of YbB$_{12}$, 240 K \cite{Iga99}.
Furthermore, the spectroscopic studies of Yb-diluted Yb$_{1-x}$Lu$_x$B$_{12}$
reported that this pseudogap feature persists in a wide range of $x$ \cite{Okamura00,Yamaguchi13}.
Therefore the large pseudogap should be attributed to the single-site Kondo effect.
On the other hand, identifying the characteristic temperature of the hybridization gap is a little more complex;
Fig.\ \ref{fig1}(d) shows the spectral intensity at $E_F$, 15, 30, and 45 meV [from Fig.\ \ref{fig1}(a)]
as functions of temperature.
In the upper panel of Fig.\ \ref{fig1}(d), one can observe a subtle kink in the $E_F$ spectral weight at
$T^\ast \sim 110$ K, resulting in a faster depletion below $T^\ast$.
Remarkably, $T^\ast$ coincides with a distinct upturn of the 15-meV spectral weight that develops into a peak
(upper panel) and an onset of development of the 30- and 45-meV spectral weight (lower panel) upon cooling.
These facts demonstrate the hybridization-gap opening at the characteristic temperature
$T^\ast \sim 110$ K.
We discuss this issue in depth later in terms of TrPES data.

% \subsection{TrPES using 1.5-eV pump, 5.9-eV probe pulses}

\begin{figure}
\begin{center}
\includegraphics[width=80mm]{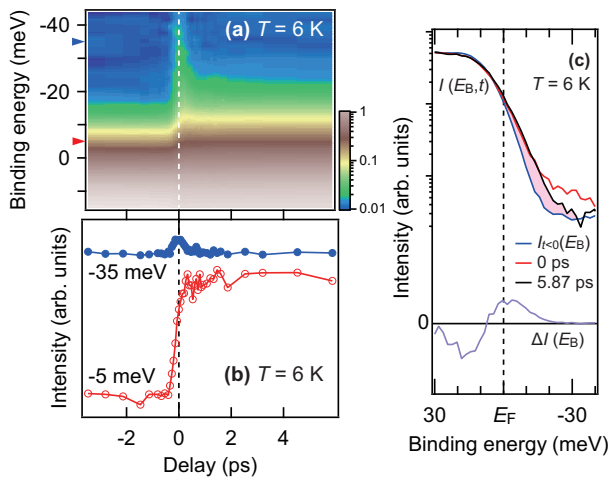}
\end{center}
\caption{
(a) TrPES intensity map at 6 K as functions of binding energy and delay time with
	a logarithmic color scale.
(b) Delay time profiles at the binding energies of $-5$ and $-35$ meV, corresponding
	to the triangles shown in panel (a).
(c) TrPES spectra recorded at 6 K displayed on a logarithmic intensity scale.
	$\Delta I(E_B)$ (see text) is also shown on a linear intensity scale.
}
\label{fig2}
\end{figure}

Figures \ref{fig2}(a) and \ref{fig2}(b) show the spectral weight in the unoccupied side in
the transient process, which is highlighted in a logarithmic TrPES intensity image [\ref{fig2}(a)] and their delay
profiles at $-5$ and $-35$ meV [\ref{fig2}(b)].
When the pump pulse $h\nu_1$ is irradiated at $t = 0$, the valence-band electrons are
photoexcited, so that the spectral intensity in the occupied side decreases and tails into the
unoccupied side.
We observed that the high-energy ($<$$-30$ meV) hot electrons relaxed into low-energy thermalized
electrons through intraband transitions within $<$1 ps, similar to the case in a typical metal Au \cite{Fann92}.
Nevertheless, it is apparent that the electron distribution is not completely recovered even
at $>$1 ps, which is clearly seen as the shaded area in Fig.\ \ref{fig2}(c).
This indicates that the recovery dynamics of the electronic system is bottlenecked.

To clarify the features in the anomalously long recovery, the difference spectra
$\Delta I(E_B)=I(E_B,\text{$t=5.87$ ps})-I_{t<0}(E_B)$ are also shown in Fig.\ \ref{fig2}(c),
where $E_B$ is the binding energy, $I(E_B,t)$ is the PES spectrum at the delay time $t$, and
$I_{t<0}(E_B)$ is the average of five spectra recorded at $-3.46 \text{ ps} \leq t \leq -1.33 \text{ ps}$
representing the spectrum before pumping.
Note, when SPV emerge at $T<T^\ast$, $I_{t<0}(E_B)$ is rigidly shifted from the spectrum recorded
without the pump; see later and the Supplemental Material in Ref.\ \cite{footnote}.
$\Delta I(E_B)$ exhibits a peak-and-valley feature throughout the transient, revealing the
accumulation of the photoexcited electrons due to the bottleneck in the recovery dynamics.
The valley bottom coincides with the 15-meV peak position observed in the valence-band
spectra (Fig.\ \ref{fig1}) that should be identified as the hybridization-gap edge in the occupied side.
The counterpart in the unoccupied side is the peak top located just above $E_F$.
The hybridization gap thus does not have an electron-hole symmetry but is asymmetric regarding $E_F$,
resulting in the gap bottom centered below $E_F$.
This is consistent with recent infrared spectroscopy which reported an indirect hybridization-gap size of
$\sim$15 meV  \cite{Okamura98,*Okamura05}.

It is noteworthy that the photoexcited electrons and holes do not recombine through
the in-gap metallic states.
Our detection of the bottlenecked electronic recovery thus has twofold implications:
(i) The scattering between the states consisting of the hybridization-gap and the in-gap states is negligibly
small, which will impose tight constraints on the origin of the in-gap states (discussed later);
(ii) TrPES can spectroscopically identify the hybridization gap as a hard charge gap that may be buried
 in the in-gap states, which strongly motivates us to investigate the $T$-dependent dynamical response
of the hybridization gap as below.

\begin{figure}
\begin{center}
\includegraphics[width=80mm]{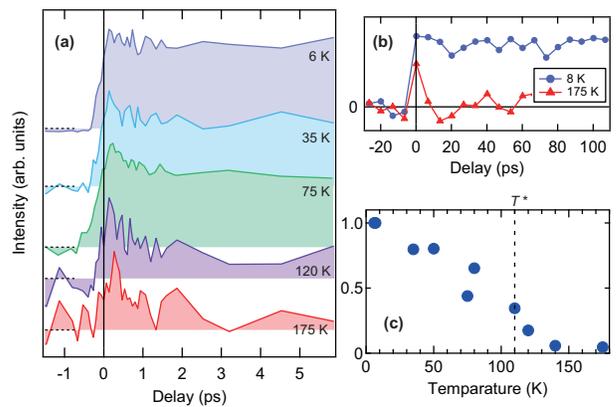}
\end{center}
\caption{
(a) Time dependence of the integrated intensities in the unoccupied region
	($-45 \text{ meV} <E_B < E_F$)
	at several temperatures. Dashed lines correspond to the zero intensity.
(b) The long range decay behaviors ($\gg$100 ps) at 8 and 175 K.
(c) Temperature dependence of the integrated intensity between 10 and 90 ps obtained from
	data shown in (a).
	The intensities were normalized to the lowest-temperature data.
	The characteristic temperature ($T^\ast$) shown in Fig.\ \ref{fig1}
	is indicated by a dashed line at 110 K.
}
\label{fig3}
\end{figure}

Now we discuss the temperature dependence of the transient response.
We show time-dependent spectra at several temperatures between 6 and 175 K in
the Supplemental Material
\footnote{See the Supplemental Material at http://dx.doi.org/10.1103/PhysRevB.92.161108 for
delay-time and temperature dependences of the near-$E_F$ spectrum as a video file.}.
Here, we found that the accumulation of excited electrons \textit{gradually}
weakened on warming and was hardly observed at $T>T^\ast \sim 110$ K.
This behavior can be explained by the fact that the photoexcited electrons rapidly go back
into the occupied state via the intraband relaxation at high $T$, indicating the
insulator-to-metal crossover with $T$.
Since the time resolution of the TrPES measurements was  $\sim$400 fs, such an intraband
rapid relaxation could not be observed clearly.
To discuss the temperature dependence of the long-lived component as a measure
of the hybridization-gap formation, the spectral weight evolutions due to the accumulated electrons
at 6--175 K are shown in Fig.\ \ref{fig3}(a).
These are defined as the integrated intensity in the region of the unoccupied side
at each temperature.
The long-lived component emerges on cooling and increasingly grows
below $\sim$100 K.
As shown in Fig.\ \ref{fig3}(b), this long-lived component does not go back to the equilibrium
state over 100 ps.
This feature can be seen more clearly in the temperature dependence of the integrated
intensity between the delay of 10 and 90 ps as shown in Fig.\ \ref{fig3}(c).
Here, one can see that the long-lived component begins to be enhanced around 100--120 K
upon cooling.
Taking into account of the laser heating of $\sim$10 K, this crossover like increase in the long-lived
component should correspond to $T^\ast \sim 110$ K determined by the ultra-high-resolution
laser-PES measurements.

\begin{figure}[t]
\begin{center}
\includegraphics[width=80mm]{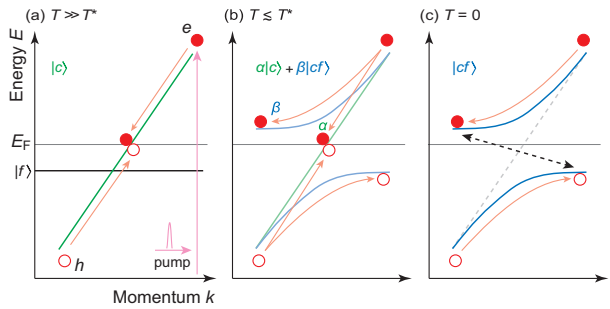}
\end{center}
\caption{
Schematic electron- ($e$-) hole ($h$) recombination processes
in (a) the metallic state ($T \gg T^\ast$), (b) the crossover state
($T\lesssim T^\ast$), and (c) the insulating ground state ($T=0$).
$|c\rangle$, $|f\rangle$, and $|cf\rangle$ correspond to the states of the conduction band,
the localized $f$ state, and the $c$-$f$ hybridized band, respectively.
}
\label{fig4}
\end{figure}

Figures \ref{fig4}(a)--\ref{fig4}(c) show schematics of the electron relaxation processes in
metallic, crossover, and insulating (ground-state) phases.
In the metallic phase ($T \gg T^\ast$), a photoexcited electron can rapidly recombine with a hole through the
intraband relaxation process.
In the ground state,  the $c$-$f$ hybridization makes a narrow and indirect gap at $E_F$.
The long-lived component is caused by the electron-hole indirect recombination across the hybridization gap,
which should be assisted by phonon or magnon scattering with a large momentum transfer
[dashed arrow in Fig.\ \ref{fig4}(c)].
A computational study on the Kondo lattice model based on the nonequilibrium dynamical
mean-field scheme also predicted such a long-lived feature in the Kondo-insulator limit (small-$J$ limit,
where $J$ is the Kondo coupling), being consistent with our observation \cite{Werner12,*Oka_pc}.
Therefore, $T^\ast$ can surely be interpreted as the characteristic temperature of the intrinsic hybridization-gap
formation in YbB$_{12}$.

Another piece of evidence for the gap formation is the SPV effect:
At $T<T^\ast$, $I_{t<0}$ exhibited a photovoltaic shift from the spectrum recorded without a pump \cite{footnote}.
SPV is observed when a charge gap exists in bulk and a surface band bending develops in the surface region,
a case often realized on semiconductor surfaces \cite{Spencer13,Ogawa13}.
SPV is also reported in recent TrPES studies on a Kondo insulator SmB$_6$ \cite{Ishida15} and
a bulk insulating topological insulator Bi$_2$Te$_2$Se \cite{Neupane15}.
The carrier dynamics at $T<T^\ast$ (Figs.\ \ref{fig2} and \ref{fig3}) is regarded as pump-induced changes occurring
in a periodic steady state realized by the irradiation of pump pulses arriving at the interval of 4 $\mu$s.

Our method, the combination of ultra-high-resolution laser PES and TrPES, has
revealed that the in-gap Fermi edge coexists with the intrinsic hybridization gap in YbB$_{12}$.
To explain the coexistence of the hybridization-gap and the in-gap states, we consider a crossover of the two
states, i.e., the nonhybridized conduction band $|c\rangle$ and the $c$-$f$ hybridized band $|cf\rangle$
[Fig.\ \ref{fig4}(b)].
Since the experiments were carried out at finite temperatures, even at the lowest one, we are practically
observing the crossover state $\alpha |c\rangle + \beta |cf\rangle$, consisting of the high-$T$ metallic state
and the insulating ground state.
Therefore, there should exist two intraband decay channels of $\alpha$ (metallic component) and
$\beta$ (insulating component) as shown in Fig.\ \ref{fig4}(b).
The gradual temperature dependence of the in-gap state [Fig.\ \ref{fig1}(d)] and the long-lived component
can be interpreted as the gradual change in the probability ratio between $|\alpha|^2$ and
$|\beta|^2$ around $T^\ast \sim 110$ K.

However, the extrapolated intensity at $E_F$ [Fig.\ \ref{fig1}(d)] towards 0 K seems to be as much as
about 30\% of the one at 225 K, indicating that the in-gap state exists even in the ground state.
Since the experimental methods that have observed pseudogap like features of the hybridization gap in YbB$_{12}$ (PES \cite{Susaki96,Susaki99,Susaki01,Takeda06,Yamaguchi13}, including this Rapid Communication, and tunneling
spectroscopy \cite{Ekino99,Batkova06}) are essentially surface sensitive probes, these measurements should
detect the surface state to some extent.
Thus we will not rule out the possibility that the observed in-gap state includes a contribution from
a metallic surface.
Such a metallic surface state in a band insulator creates much interest in topological insulators \cite{Moore10,*Hasan10,*Ando13}.
Recently, there have been impressive discussions on Kondo insulators SmB$_6$, YbB$_6$, and YbB$_{12}$
as possible three-dimensional topological insulators by both theory \cite{Dzero10,*Takimoto11,*Lu13,*Weng14}
and experiment \cite{Xu13,*Jiang13,*Wolgast13,*Kim14,*Zhang13,*Neupane13,*Xia14,*Xu14}.
Although the in-gap states can originate from a fractured surface quality (e.g., a surface defect state),
the observed Fermi-edge state coexisting with the insulating feature in the TrPES data may meet the criteria
of a topological Kondo insulator.

% \section{Summary}

To summarize, the low-energy electronic structure and its transient properties
of the Kondo insulator YbB$_{12}$ were investigated using ultrahigh-resolution PES and TrPES.
In the $T$-dependent laser-PES spectra, we found two different (pseudo)gaps
with sizes of 25 and 15 meV which are attributed to the single-site effect and the insulating
hybridization-gap opening, respectively. 
The characteristic temperature $T^\ast$ was determined to be $\sim$110 K where the
hybridization gap begins to open, although the Fermi edge remains as the in-gap state
event at the lowest $T$.
In TrPES measurements, we found that the long-lived ($\gg$100 ps) component gradually develops
upon cooling through $T^\ast$, which is taken as a signature of the hybridization-gap evolution.
Thus we experimentally determined the characteristic temperature
 $T^\ast \sim 110$ K as the metal-to-insulator crossover in YbB$_{12}$.

% \begin{acknowledgments}
The authors thank T.\ Oka, P.\ Werner, K. Tsunetsugu, and K. Ueda for useful discussions.
This work was supported by MEXT KAKENHI Grant No.\ 20102004 and JSPS KAKENHI
Grants No.\ 23540413, No.\ 23840039, No.\ 23740256, No.\ 25400378, No.\ 25420174, No.\ 26400321, and No.\ 26800165.
JSPS supported this research also through the FIRST Program, initiated by the Council for
Science and Technology Policy.

\bibliographystyle{apsrev4-1}
\bibliography{references}

\end{document}